\documentclass[12pt]{scrartcl}
\usepackage[utf8]{inputenc}
\usepackage{graphicx}
\usepackage{array} 
\usepackage{subfig}
\usepackage{slashed}
\usepackage{amsmath}
\usepackage{amsfonts}
\usepackage{amssymb}
\usepackage{cite}
\usepackage{epsfig}
\usepackage{comment}
\usepackage{hyperref}
\usepackage{tikz}
\usepackage{feynmp}
\usepackage{cancel}
\usepackage{mathrsfs}
\usepackage{mathtools}
\usepackage{fourier}

\numberwithin{equation}{section}

 \def\p{\partial}

\def\0{{(0)}}
\def\1{{(1)}}
\def\2{{(2)}}

\def\<{\langle }
\def\>{\rangle }

\newcommand{\bea}{\begin{eqnarray}}
\newcommand{\eea}{\end{eqnarray}}
\newcommand{\be}{\begin{equation}}
\newcommand{\ee}{\end{equation}}
\newcommand{\ba}{\begin{align}}
\newcommand{\ea}{\end{align}}

\renewcommand{\epsilon}{\varepsilon}

   \makeatletter
  \let\over=\@@over \let\overwithdelims=\@@overwithdelims
  \let\atop=\@@atop \let\atopwithdelims=\@@atopwithdelims
  \let\above=\@@above \let\abovewithdelims=\@@abovewithdelims
\renewcommand\section{\@startsection {section}{1}{\z@}%
                                   {-3.5ex \@plus -1ex \@minus -.2ex}%nn
                                   {2.3ex \@plus.2ex}%
                                   {\normalfont\large\bfseries}}

\renewcommand\subsection{\@startsection{subsection}{2}{\z@}%
                                     {-3.25ex\@plus -1ex \@minus -.2ex}%
                                     {1.5ex \@plus .2ex}%
                                     {\normalfont\bfseries}}

\newcommand{\beq}{\begin{equation}}
\newcommand{\eeq}{\end{equation}}
\newcommand{\beqa}{\begin{eqnarray}}
\newcommand{\eeqa}{\end{eqnarray}}
\newcommand{\beqar}{\begin{eqnarray*}}

\def\[{\big[}
\def\]{\big]}

\newcommand{\bd}[1]{\begin{fmffile}{#1}\begin{fmfgraph*}}
\newcommand{\ed}{\end{fmfgraph*}\end{fmffile}}

\begin{document}
\begin{titlepage}
\unitlength = 1mm
\ \\
\vskip 3cm
\begin{center}

{ \LARGE {\textsc{On massless dyadic forms and no minimal coupling theorem}}}

\vspace{0.8cm}
Alexis Kassiteridis\footnote{A.Kassiteridis@physik.uni-muenchen.de}

\vspace{1cm}

\begin{small}
{\it Arnold Sommerfeld Center for Theoretical Physics, Ludwig-Maximilians-Universit\"at\\
Theresienstraße 37, D-80333 M\"unchen, Germany}
\end{small}

\begin{abstract}
We use the spinor helicity formalism in order to derive the dyadic forms for massless fields of various spins. We also give an iterated form of this approach in case higher spin theories are under study. This reduces calculations at hard and soft scattering problems in gauge theories drastically. Finally, we state and prove a theorem of gauge symmetry violation in the presence of minimal coupling with light in higher spin theories ($j>1/2$). 
 \end{abstract}

\vspace{1.0cm}

\end{center}

\end{titlepage}

\pagestyle{empty}
\pagestyle{plain}

\def\vx{{\vec x}}
\def\p{\partial}
\def\po{$\cal P_O$}

\pagenumbering{arabic}

\tableofcontents
\vspace{1cm}

\section{Introduction}
These notes are organized as follows. In section \ref{section definitions} we present our conventions and definitions of the Dirac wave functions and present useful identities. In section \ref{section massless 1} we compute the dyadic form of a massless spin-1 field introducing a useful representation of its polarization vectors. In section \ref{section massless 2} we continue our discussion about dyadic forms considering the massless graviton and in section \ref{section massless 3/2} the Rarita-Schwinger field \cite{raritaschwinger}; our aim is find an iteration for higher spin fields and which is sketched in section \ref{section higher spins}. The no minimal coupling theorem is presented for the first time with a proof in section \ref{section theorem}. Finally, in section \ref{section conclusion} we conclude.
%\vspace{0.8cm}

\section{Mathematical framework - Dirac wave functions} \label{section definitions}
First we define the wave functions with negative helicity of the free massless\footnote{In abuse of language we will refer to helicity eigenstates and we will also mean chirality eigenstates, since in the massless case they coincide.} Dirac field $\Psi(x)$ using the one particle state with three-momentum $\textbf{p}$ and spin $s$
\begin{equation} \label{Spinor definition L}
(\Omega ,   (P_L \Psi(0))_{{\alpha}} \sum_s 1_{\textbf{p}+s} ) := \ _{{\alpha}} p] \quad \text{with} \quad \gamma_5 p] = - p].
\end{equation}
Analogously for the positve helicity particle eigenstate we have
\begin{equation} \label{Spinor definition R}
(\Omega , (P_R \Psi (0))^{\dot{\alpha}} \sum_s 1_{\textbf{p}+s} ) := \ ^{\dot{\alpha}} p\rangle \quad \text{with} \quad \gamma_5 p\rangle = + p\rangle.
\end{equation}
Here $P_{L/R}$ are the usual helicity projectors on the spinor space. \\
For the anti-particle case, one obtains similar relations up to a free constant phase which we set to zero. Therefore, we have following universal mapping:
\begin{align*}
&\text{For left-handed particles or right-handed anti-particles}: \quad &\ _{{\alpha}} p]  \\
&\text{For right-handed particles or left-handed anti-particles}:  \quad &\ ^{\dot{\alpha}} p\rangle 
\end{align*}
In order to complete the presentation of the wave-functions we give also the Dirac adjoint ones:
\begin{align*}
&\text{For left-handed particles or right-handed anti-particles}: \quad &\langle p_{\dot{\alpha}}   \\
&\text{For right-handed particles or left-handed anti-particles}: \quad &[ p^{{\alpha}}
\end{align*}
The reader can easily test the following relations for massless helicity spinors
\begin{equation*}
\langle p p ] =  [p p \rangle = \langle p \slashed{p} = \slashed{p} p \rangle = [p \slashed{p} = \slashed{p} p ]  =0
\end{equation*}
together with
\begin{equation*}
\langle i j \rangle^* = \langle i j \rangle^\dagger = [ j i ]
\end{equation*}
where we used the abbreviation $p_i \equiv i$. \\
The spinor projectors take the following form
\begin{equation*}
p\rangle [ p = \slashed{p} P_L \quad \text{and} \quad p] \langle p = P_L \slashed{p}
\end{equation*}
and span the spinor space
\begin{equation} \label{Dirac projector}
p] \langle p + p\rangle [ p = [P_L , \slashed{p} ]_+ = \slashed{p}.
\end{equation}
This is nothing else than the sum over all polarizations of the massless Dirac field and the corresponding dyadic is then given by
\begin{equation} \label{Dirac dyadic}
D^{ab}_\text{Dirac} (p) =\   _{{\alpha}}  p] \langle p_{\dot{\beta}} +\ ^{\dot{\alpha}}p\rangle [ p ^{\beta} = \slashed{p}^{ab} .
\end{equation}
We can use the above projectors to compute the following probabilities\footnote{This is not normalized to 1 for convenience.}
\begin{equation*}
\vert \langle i j \rangle \vert^2 = \vert [ i j] \vert^2 = 2\ i\cdot j
\end{equation*}
using elementary trace identities of the $\gamma$-matrices.\\
Applying the definition of charge conjugation for Dirac wave functions one gets the familiar anti-symmetry property, namely
\begin{equation*}
\langle i j \rangle = - \langle j i \rangle \quad \text{and} \quad [ij] = -[ji].
\end{equation*}
Obviously $\langle i i \rangle = [ii]= 0$.\\
Therefore, it is useful to express the amplitudes as 
\begin{equation*}
\langle i j \rangle = \sqrt{2 \ i\cdot j} \ \mathrm{e}^{\mathrm{i}\phi_{ij}} 
\end{equation*}
with $\phi_{ij} \in \mathbb{R}$ and similar for the other helicity pairs.\\
Closing this section we notice two important relations:
\begin{enumerate}
\item [(i)] We reformulate the Fierz identity
\begin{equation} \label{Fierz identity}
\langle i \mu j ] \langle k \mu l ] = 2 \langle \mathrm{i} k \rangle [l j] \quad \text{and} \quad \langle i \mu j ]  [k \mu l \rangle = 2 \langle i l \rangle [kj]
\end{equation}
 with $\gamma^{\mu \updownarrow} \equiv \mu $ and repeated Greek indexes ignite appropriate contractions.
\item [(ii)] The Schouten identities read
\begin{equation} 
 \langle i j \rangle \langle k l \rangle + \langle j k \rangle \langle i l \rangle + \langle k i \rangle \langle j l \rangle =0\ .
 \end{equation} 
\end{enumerate}

%\vspace{1cm}
 
 \subsection{Application: crossing symmetrical $\beta$-decay}
As a pedagogical application we consider the following scattering in SM-framework at low energies \cite{fermi} but still not trivial to consider massless matter: 
\begin{equation*}
 d (1)\  \nu_e (2)  \rightarrow u (3) \ e(4).
 \end{equation*} 
The quantum mechanical amplitude at the lowest order of weak-coupling\footnote{Here the weak-coupling $G_w$ is defined as the Fermi $\beta$-decay constant $G_F$ \cite{fermi} times the Cabibbo mixing parameter $\cos \theta_C$.} $G_w$ is given by
\begin{equation*}
 \mathcal{A}(1^-2^-\rightarrow 3^-4^-) = -\mathrm{i} 2\sqrt{2}G_w\  \langle 3 \mu 1]\  \langle 4 \mu 2].
\end{equation*}
Using identity (\ref{Fierz identity}) we get
\begin{equation*}
\mathcal{A}(1^-2^-\rightarrow 3^-4^-) = -\mathrm{i}  4\sqrt{2}G_w\  \langle 3 4 \rangle\  [2  1]
\end{equation*}
and therefore the modulus squared takes automatically the following form using the spinor identities stated above
\begin{equation*}
\vert  \mathcal{A} \vert^2 = 128 G^2_w \ 3\cdot 4 \ 2 \cdot 1.
\end{equation*}
The job is now done. This gives rise to the familiar total cross section
\begin{equation*}
\sigma_\text{tot} = \frac{G^2_w}{{\pi}}  s
\end{equation*}
where $s$ is the CMS energy of the scattering. \\
One sees the power of this formalism in computing processes at the level of the quantum amplitude and not at the level of probability as the older methods. An enhancement of efficiency and speed of calculation is the major contribution of this method.
%\vspace{1cm}

\section{Massless dyadic forms} 
\subsection{Massless spin-1 field} \label{section massless 1}
We now turn our attention to real massless spin-1 fields, with ultimate goal to derive the corresponding dyadic form.\\
It is true that one can combine two spinors to construct a vector \cite{schwinger}, therefore, we define taking $p \cdot p=0$ the following Lorentz vectors
\begin{equation} \label{Vector definition}
\epsilon^\mu_-(p;\bar{p}) :=  N_-^{-1} [ p \mu \bar{p} \rangle \quad \text{and} \quad \epsilon^\mu_+(p;\bar{p}) :=  N_+^{-1} \langle p \mu \bar{p} ] 
\end{equation}
where $\bar{p} \equiv \mathcal{P}(p) $ is the backwards momentum with $p \cdot \bar{p}\neq 0$ and $\bar{p} \cdot \bar{p}=0$. Furthermore $N_\pm \in \mathbb{C}$.\\
One immediately sees that 
\begin{equation} \label{Vector relations1}
p \cdot \epsilon_\pm(p;\bar{p}) = \bar{p} \cdot \epsilon_\pm(p;\bar{p}) = \epsilon_+(p;\bar{p}) \cdot \epsilon_-(p;\bar{p}) =0 .
\end{equation}
In addition, we fix the normalization constant. We demand
\begin{equation} \label{Vector relations2}
\epsilon_\pm(p;\bar{p}) \cdot \epsilon_\pm^*(p;\bar{p}) = -1
\end{equation}
and we get 
\begin{equation*}
N \equiv N_\pm = \sqrt{4 \ p \cdot \bar{p}} .
\end{equation*}
The relations (\ref{Vector relations1}) and (\ref{Vector relations2}) are nothing more than the constraints on the polarization vectors of a real massless spin-1 field. This gives rise to the following dyadic form
\begin{align} \label{Vector dyadic}
D^{\mu\nu}_\text{spin-1} (k; \bar{k}) := & \sum_{s\in \{\pm\}}  \epsilon^\mu_s(k;\bar{k}) \ \epsilon^{*\nu}_s(k;\bar{k}) \nonumber \\ 
= & N^{-2}[ \bar{k} \mu k \rangle [ k \nu \bar{k} \rangle + N^{-2} \langle \bar{k} \mu k ]  \langle k \nu \bar{k} ] \nonumber \\
 = & -g^{\mu\nu} + 2 \frac{k^{(\mu} \bar{k}^{\nu )} }{k \cdot \bar{k}}.
\end{align}
We obtained this elegant result using elementary trace identities of the $\gamma$-matrices.\\
It is instructive to notice that contracting the dyadic $D^{\mu\nu}_\text{spin-1} (k; \bar{k})$ with conserved currents\footnote{The conservation of $J^\mu$ requires $p \cdot J(p) =0$ but not when contracted with $\bar{p}^\mu$. } the pure gauge term vanishes \cite{schwinger}. In the following discussion we will always ignore such terms.
%\vspace{1cm}

\subsection{Massless spin-2 field} \label{section massless 2}
In this section we derive the dyadic forms of the massless graviton and the Rarita-Schwinger field.\\
For the massless spin-2 symmetric tensor field under some convenient gauge fixing only two degrees of freedom propagate (physical field). Therefore, the following decomposition holds \cite{schwinger}
\begin{equation*}
\vert \text{graviton} \rangle = \vert \text{photon} \rangle \otimes \vert \text{photon} \rangle .
\end{equation*}
We write the corresponding polarization tensor $\epsilon^{\mu\nu}_s(k,\bar{k})$ as 
\begin{equation}
\epsilon^{\mu\nu}_{\pm 2}(k;\bar{k}) = \epsilon^\mu_\pm(k;\bar{k})\ \epsilon^\nu_\pm(k;\bar{k}) .
\end{equation}
The above mapping allows us to compute the dyadic of the field summing over all physical polarizations,
\begin{align} \label{Tensor2 dyadic}
D^{\mu\nu, \kappa \lambda }_\text{spin-2} (k; \bar{k}) := & \sum_{s\in \{\pm 2\}} \epsilon^{\mu\nu}_s(k;\bar{k})\  \epsilon^{*\kappa \lambda}_s(k;\bar{k}) \nonumber \\
 = & \sum_{s\in \{\pm\}} \epsilon^{\mu}_s(k;\bar{k})\ \epsilon^{\nu}_s(k;\bar{k})\ \epsilon^{*\kappa}_s(k;\bar{k})\  \epsilon^{*\lambda}_s(k;\bar{k})   \nonumber \\ 
 = & \tfrac{1}{2} \sigma^{\mu\kappa\nu \lambda}   + \tfrac{1}{ k \cdot \bar{k}} \big[ g^{\nu \lambda} (\bar{k}^{[\kappa} k^{\mu ]}   - k^\mu \bar{k}^\kappa) + g^{\kappa \nu} (\bar{k}^{[\mu} k^{\lambda ]} - k^\lambda \bar{k}^\mu) \nonumber\\ 
 &+ g^{\kappa \lambda} \bar{k}^{(\nu} k^{\mu )}   - g^{\mu \kappa } \bar{k}^{(\nu} k^{\lambda )}  - g^{\mu \nu} (\bar{k}^{[\kappa} k^{\lambda ]} - k^\lambda\bar{k}^\kappa) - g^{\mu \lambda } \bar{k}^{(\nu} k^{\kappa )}  \big] \nonumber\\
 & + \tfrac{2}{ (k \cdot \bar{k})^2} \big[ k^\mu \bar{k}^\kappa \bar{k}^{(\nu} k^{\lambda)} + k^\lambda \bar{k}^\mu \bar{k}^{(\nu} k^{\kappa)}  - k^\lambda \bar{k}^\kappa \bar{k}^{(\mu} k^{\nu)} \big] \nonumber \\
 = & \tfrac{1}{2} \sigma^{\mu\kappa\nu \lambda} + \text{pure gauge}
\end{align}
Again we computed this dyadic using elementary trace identities of the $\gamma$-matrices.\\
The pure gauge term is of no great interest since the gravitational source $T_{\mu\nu}$ is manifestly conserved.
%\vspace{1cm}

\subsection{Massless spin-3/2 field} \label{section massless 3/2}
In the case of a massless Rarita-Schwinger field we follow the same problematic: we decompose the field in a vector and spinor.  It can be easily shown that after applying the physical state condition $\gamma^\mu \Psi_\mu=0$ \cite{raritaschwinger}, the maximal/minimal helicities survive.\\
In other words, the $+3/2$ polarization tensor reads
\begin{equation} \label{correct helicity}
\epsilon_{+3/2}^{\mu, \dot{\alpha}} (p; \bar{p})  = N^{-1} \langle p \mu \bar{p} ] \ ^{\dot{\alpha}} p \rangle
\end{equation}
and similarly the $-3/2$ polarization tensor is given by
\begin{equation*}
\epsilon_{-3/2}^{\mu, \alpha \downarrow} (p; \bar{p})  = N^{-1} [ p \mu \bar{p} \rangle \ _\alpha p ].
\end{equation*}
We are now ready to compute the dyadic form\footnote{We do not really keep track of the spinor labels, since they exchange positions in the two terms.}
\begin{align} \label{Tensor3/2 dyadic}
D^{\mu \nu ,ab}_\text{spin-3/2} (p; \bar{p}) :=  & \sum_{s\in \{\pm 3/2\}} \left(\epsilon_{s}^{\mu,{\dot{\alpha}}\updownarrow} (p; \bar{p})\ \bar{\epsilon}_{s}^{\nu, \beta \updownarrow} (p; \bar{p}) \right)^{ab} \nonumber \\
= &  \tfrac{1}{2}\ ^{a}\gamma^\nu \slashed{p} \gamma^{\mu, b} + \text{pure gauge}
\end{align}
again using elementary identities of the $\gamma$-matrices. The pure gauge term is manifestly zero after contracting it with a conserved fermionic current $\eta_{\nu}^b$.\\
Here, it is important to notice that such currents in an interacting theory emerge from a supersymmetric source \cite{Das:1976ct} and a minimal coupling does not fulfil this condition \cite{weinbergwitten}, \cite{Berends:1979kg}. We will prove a more general version of this statement in section \ref{section theorem}.\\
For the sake of completeness we present the full version of the dyadic form of massless Rarita-Schwinger field:
\begin{align*}
D^{\mu \nu ,ab}_\text{spin-3/2} (p; \bar{p}) & = \frac{^a \slashed{p}\gamma^\nu \slashed{\bar{p}}\gamma^\mu \slashed{p}^b}{4 p\cdot \bar{p}} \\
& = \tfrac{1}{2} \left(\ ^a\gamma^{\nu} \slashed{p} \gamma^{\mu,b} +2\ ^a\slashed{p}^b p^\mu \bar{p}^\nu -\  ^a\gamma^{(\nu,b}p^{\mu )} + \frac{^a\slashed{\bar{p}}\slashed{p} \gamma^{\nu,b} p^\mu +\ ^a\slashed{p}\slashed{\bar{p}} \gamma^{\mu,b} p^\nu}{ p\cdot \bar{p}} \right).
\end{align*}
By inspection we see that $D^{\mu \nu ,ab}_\text{spin-3/2}  $ is the actual projector respecting the algebraic gauge fixing and the equations of motion and rejects the longitudinal and $\pm1/2$ degrees of freedom.
%\vspace{1cm}

\subsection{Higher spins} \label{section higher spins}
At this point it should be clear how the algorithm looks in order to calculate dyadic forms of a massless higher spin field. We iterate the decomposition in $n$ vector currents (\ref{Vector definition}) and one spinor (\ref{Spinor definition L}), (\ref{Spinor definition R}); then we construct polarization tensors for massless $n+1/2$ spin fields. I.e.
\begin{equation} \label{Higher spin iterate}
\epsilon_{\pm(n+1/2)}^{\mu_1...\mu_n, \alpha \updownarrow} (p ; \bar{p}) = \epsilon^{\mu_1}_\pm(p;\bar{p})\ ...\ \epsilon^{\mu_n}_\pm(p;\bar{p}) \ (\Omega ,   (P_{R/L} \Psi(0))^{\alpha \updownarrow} \sum_s 1_{\textbf{p}+s} )
\end{equation}
for the maximal helicity states\footnote{Here we used the formal decomposition for arbitrary polarizations, keeping track of the correct phases (\ref{correct helicity}).}.\\
Hence\footnote{We assume that all greek indexes of the same family are symmetrized properly.} we have for $n \in \mathbb{N}$ an iteration formula\footnote{We understand under $\rm{Tr}\left(\prod^0\right) =1$ that no spin is present.}
\begin{align} \nonumber
D^{\mu_1...\mu_n, \nu_1...\nu_n}_{\text{spin-}n}(k;\bar{k}) = & \tfrac{1}{(4 k\cdot \bar{k})^n}\text{Tr}\left[\prod^n_j \left( \slashed{\bar{k}}\mu_j \slashed{k} \nu_j\right)\right] \quad \text{and} \\
D^{\mu_1...\mu_n \nu_1...\nu_n,ab}_{\text{spin-}n+1/2}(p;\bar{p}) = & \tfrac{1}{(4 p\cdot \bar{p})^n} \ ^a\slashed{p}  \prod_j^n \left( \gamma^{\nu_j} \slashed{\bar{p}}\gamma^{\mu_j} \slashed{p}\right)\ ^b
\end{align}
We see at once the power of this formalism: without knowing the explicit form of the underlying Lagrangian and without even trying to find a suitable gauge constraint, we only use the desired statistics and helicity values of the field. Then the construction of the corresponding dyadic form reduces to an elementary exercise of Clifford algebra gymnastics.\\
Such higher spin fields do not allow mediation of forces of inverse square law and if they exist they will not introduce any long range effects \cite{sweinberg} nor they admit conserved charges in an interacting theory \cite{weinbergwitten}.

%\vspace{1cm}

\section{No minimal coupling theorem} \label{section theorem}
In this section, continuing the previous discussion from section \ref{section massless 3/2}, we present a theorem about the incompatibility of minimal coupling and bosonic or fermionic gauge symmetries.\\\\
\textit{Theorem: If a minimal coupling (to the photon) exists in a field theory, where the massless field $\Psi$ carries at least one Lorentz-vector label, then its gauge symmetry (bosonic or fermionic) $\mathcal{G}$ is broken.}\\\\
In the free theory there exists a conserved source $\eta$ such that $\partial \eta = 0$, with $\eta \propto \langle 1_{\textbf{p} } \vert 0_- \rangle^\eta$ \cite{schwinger}.  Therefore, $\exists \mathcal{G}: \ \Psi {\longmapsto} \Psi + \partial \chi$ and the vacuum persistence amplitude, $\mathrm{e}^{\rm{i}W[\eta]}$, is intact due to the invariance of the action $I[\Psi]$ under $\mathcal{G}$.\\
According to the hypothesis, there exists a minimal coupling to the photon, namely
\begin{equation} \label{theorem current}
A\cdot J \quad \text{where} \quad  J(x)= \underbrace{\rm{-i} \frac{\delta I[\Psi]}{\delta \partial\Psi(x)} \delta \Psi(x) }_{\text{Noether current}} + P[\Psi,A](x) .
\end{equation}
In other words, $P[\Psi,A](x)$ is $J(x)$ minus the conserved Noether current of the global symmetry and therefore a polynomial of fields and their derivatives.\\
Without any loss of generality, we define the corresponding new source $\eta'$ of interacting field $\Psi$ as follows:
\begin{equation} \nonumber
\eta' := \eta + A \cdot \frac{\delta \mathcal{J}}{\delta \Psi}
\end{equation}
with $ \mathcal{J}[\Psi, A](t) \equiv \int \mathrm{d^3x} J(x)$.\\
If $\mathcal{G}$ is still unbroken then $\partial \eta'=0$, hence
\begin{equation*}
\partial_\mu \eta'^\mu = \underbrace{\partial_\mu \eta^\mu}_{=0} + \partial_\mu A_\nu \frac{\delta \mathcal{J}^\nu}{\delta \Psi_\mu} + A_\nu \partial_\mu \frac{\delta \mathcal{J}^\nu}{\delta \Psi_\mu} .
\end{equation*}
Here following J.Schwinger \cite{schwinger}, $\eta$ is simply the per definition conserved field-generating source of the free theory.\\
If $n$ is the maximum power of the minimal coupling constant in $I[\Psi,A]$, we construct the following term:
\begin{equation} \label{theorem g definition of eta'}
\lim_{g\rightarrow  \infty} \frac{1}{g^n}  \eta' := \eta'^{\#}
\end{equation}
which admits a four-current, $\mathcal{J}^{\#}$, free of spacetime derivatives of the fields\footnote{We are allowed to use the e.o.m. of $A$ but not of $\Psi$, since $\eta'$ is the generating source of $\Psi$.}. \\
We immediately see that no $\mathcal{J}^{\#}$, besides the trivial one $\mathcal{J}^{\#}=0$, fulfils this constraint under the assumption that $\mathcal{G}$ gauge symmetry is still present, since $\mathcal{J}^{\#}[\Psi, A](t)$ is a functional of the total field space.
% The divergence of the interacting source reads
%\begin{equation}
%\partial_\nu \eta'^{\# \nu} = \partial_\nu A_\mu \frac{\delta \mathcal{J}^{\#\mu}}{\delta \Psi_\nu}  + A_\mu \frac{\delta^2 \mathcal{J}^{\#\mu}}{\delta \Psi_\nu \delta A_%\lambda} \partial_\nu A_\lambda + 2 A_\mu \frac{\delta^2 \mathcal{J}^{\#\mu}}{\delta \Psi_\nu \delta \Psi_\lambda} \partial_{(\nu} \Psi_{\lambda)}.
%\end{equation}
%Immediately we obtain the vector-differential constraint on $\mathcal{J}^{\#}$ under the assumption that $\mathcal{G}$ gauge symmetry is still present:
%\begin{equation} \label{theorem constraint}
 %\frac{\delta \mathcal{J}^{\#\kappa}}{\delta \Psi_\nu}  + A_\mu \frac{\delta^2 \mathcal{J}^{\#\mu}}{\delta \Psi_\nu \delta A_\kappa} =0 .
 %\end{equation} 
%Since only positive powers in the fields are allowed in $\mathcal{J}^{\#}$, equation (\ref{theorem constraint}) has no solution, besides the trivial one, $\mathcal{J}^{\#}=0$.\\
The iteration is obvious; we proceed to the previous order in the coupling constant $\mathcal{O}(g^{n-1})$, taking the corresponding limit introduced in definition (\ref{theorem g definition of eta'}); now this is the term, where spacetime derivatives of the fields are absent. Again the underlying $\mathcal{J}^{\#}$ fulfils the divergence equation and vanishes. This goes all the way down to the $\propto g$-term, where by the hypothesis $\mathcal{J}^{\#} \neq 0$ (existence of minimal coupling), which leaves $\partial_\mu \eta'^\mu \neq 0$. This means that if a minimal coupling exists, then the internal $\mathcal{G}$ symmetry is absent, which concludes the proof. %$\hspace{5.25cm} \Box$
\\\\ 
This theorem has severe consequences to the interactions between light and massless bosonic or fermionic particles with $j>1/2$. We notice that during the interaction, the e.o.m. of the field are inconsistent, namely
\begin{equation*}
\mathcal{K} \Psi = - \eta' 
\end{equation*}
where $\mathcal{K}$ is the kinetic operator of the theory, makes sense only in the limit $g \rightarrow 0$.\\
In amplitudes, where $\Psi$ participates, one cannot simply substitute the Lorentzian part of the polarization vector (\ref{Higher spin iterate}), $\epsilon^{\mu_1}_\pm(p;\bar{p})\ ...\ \epsilon^{\mu_n}_\pm(p;\bar{p})$, by the on-shell momentum $p^{\mu_i}$, in order to get zero. In other words, $\bar{p}$ is fixed by the definition of the dyadic form and does not serve as a non-perpendicular auxiliary momentum in the scattering, as in the case of $gg \rightarrow gg$ process \cite{Dixon:1996wi}. This problem is already known for the Rarita-Schwinger \cite{PhysRev.186.1337}, \cite{Das:1976ct} and the spin-$5/2$ field case \cite{Berends:1979kg} and here we took the effort to confirm and prove it in general for any $j>1/2$ fields.\\
The non-existence of a minimal coupling to photon\footnote{A possible cure to this problem, at least in the Rarita-Schwinger case, is  to upgrade the internal fermionic symmetry to a generalized covariant transformation $\Psi \rightarrow \Psi + D \chi$ as described in \cite{Adler:2015hna} inspired by the super-transformations \cite{Das:1976ct}. This makes the action invariant up to a total derivative, if one imposes initial value secondary constraints.} does not forbid a gauge invariant effective interaction of neutral charged fermions; for example in the case of spin-$3/2$ field  of the type $\bar{\psi} F^{\mu\nu} \Psi_{\mu\nu}$, where the field strengths are protecting the bosonic and fermionic gauge invariance of the action functional $I[\Psi, \psi, A]$. One can easily check that in this theory the divergences of $\eta'$ and $J$ (the sources of the spin-$3/2$ and the photon field respectively) are identically zero.

%\vspace{1cm}

%\vspace{1cm}

\section{Conclusion} \label{section conclusion}
The spinor helicity formalism allows us to work right at the level of quantum mechanical amplitudes without involving squaring the scattering amplitudes and summing analytically over all spin values of the physical legs and in the end expressing the result in Lorentz invariant quantities. This approach provides an enormous advantage over the traditional methods of computing and studying hard and soft processes in massless gauge theories, reducing the average terms that one should compute drastically.\\
The power of this formalism does not end here. Besides the computation of hard and soft scattering amplitudes, we used spinor helicity eigenstates to derive the dyadic forms of massless fields with arbitrary spin, minimizing the time cost and sparing the inversion of canonical kinetic terms. These forms appear on the numerator of gauge-fixed Feynman two point functions. We also gave an iterating formula (\ref{Higher spin iterate}) for higher spin fields and stated and proved an important theorem of gauge symmetry violation in the presence of minimal coupling with light in higher spin theories ($j>1/2$), which makes charged higher spin interactions with light non-trivial in the sense of coupling.

%\vspace{1cm}

\section{Acknowledgments}
We would like to thank Stefan Hofmann, Sebastian Konopka and Oscar Cata for the long and useful discussions about different aspects of this project. We also thank Frederik Lauf, Maximilian K\"ogler and Ottavia Balducci for suggesting various improvements of these notes. This work was supported by TRR 33.

\end{document}